\documentstyle[preprint,prd,aps,eqsecnum]{revtex}
\begin{document}
\preprint{hep-th/9608156\hspace{-28.6mm}\raisebox{2.4ex}{CCNY-HEP-96/9}
          \hspace{-32.9mm}\raisebox{-2.4ex}{August 1996}}
\title{Recursion Rules for Scattering Amplitudes in \\
Non-Abelian Gauge Theories}
\author{Chanju Kim\thanks{Permanent address: Physics Department, Seoul
National University, Seoul, 151-742, Korea.} and V.P. Nair\thanks{e-mail:
cjkim@scisun.sci.ccny.cuny.edu, vpn@ajanta.sci.ccny.cuny.edu}}
\address{Physics Department\\
         City College of the City University of New York\\
         New York, New York 10031}
\maketitle
\begin{abstract}
We present a functional derivation of recursion rules for scattering
amplitudes in a non-Abelian gauge theory
in a form valid to arbitrary loop order. The tree-level
and one-loop recursion rules are explicitly displayed. 
\end{abstract}
\pacs{PACS number}
\section{INTRODUCTION}
It is by now a well-appreciated fact that non-Abelian gauge theories
display many interesting and beautiful properties which are, no doubt,
indicative of their deeply geometrical nature and symmetry. Already at
the level of perturbation theory, many of the scattering amplitudes, in
Quantum Chromodynamics (QCD) for example, have a simple and elegant
form, although large numbers of Feynman diagrams have to be summed up
to arrive at these \cite{PT}.
This feature was originally understood in terms of calculational
techniques based on string theory as well as recursion
rules \cite{BG}-\cite{BDK}. (A field theoretic understanding of the 
string-based techniques has also emerged \cite{Strassler}.)
There are also indications that there are integrable field theories
hidden in non-Abelian gauge theories which describe some aspects
of these theories, such as scattering for certain specific choices of
helicities as well as scattering in the Regge regime 
\cite{Lipatov}-\cite{Cangemi}.
Given these features, it is clear that the theoretical exploration of the
structure of even the perturbative scattering matrix can be quite useful.

Within a field-theoretic approach, recursion rules for scattering
amplitudes have been very useful in understanding color factorization
and many other properties. These rules were originally derived for 
tree-level amplitudes using diagrammatic analyses by Berends and Giele 
and have later been extended upto the one-loop level \cite{MYD}-\cite{M}.
Elaboration and extension of this technique, we feel, can be
very fruitful. In this letter, we present a derivation of the
recursion rules within a functional formalism without reliance
on diagrammatic analysis and in a way valid for any field theory.
{}From the basic equations, recursion rules valid upto arbitrary
loop order can be obtained, although at the expense of increasing
algebraic complexity. We explicitly display the tree-level and
one-loop recursion rules. Renormalization constants, which one
must consider beyond the tree-level are also easily incorporated
in a functional derivation. 
\section{THE S-MATRIX FUNCTIONAL}
We start by considering a scalar field theory with action of the form
\begin{equation}
S[\phi]=S_0[\phi]+ S_{int}[\phi]\,,
\end{equation}
where $S_0[\phi]$ is the free action of the form
$ S_0[\varphi]=\int\frac{1}{2}\phi K\phi$
and $S_{int}$ is the interaction part of the action. Specifically, $K$ can be 
taken to be of the form $K= -Z_2 (\partial^2 +m^2)$. The functional for the
S-matrix elements can be written as \cite{Faddeev}
\begin{equation}\label{2}
{\cal F}[\varphi]=e^F~\int [d \phi ]\ e^{iS_0[\phi]+iS_{int}(\phi+\varphi)}\,, 
\end{equation}
where $\varphi(x)$ can be expanded as
\begin{equation}
\varphi(x)=\sum_k a_ku_k(x)+a_k^*u_k^*(x)\,.
\end{equation}
$u_k(x), u_k^*(x)$ are the one-particle wave functions and are solutions
of the free field equation $K\varphi=0$. $F$ is given by 
$\sum_ka_k^*a_k$ and is
introduced in (\ref{2}) to give the matrix elements for subscattering
processes where some of the particles fly by unscattered. The matrix element
for a process where particles of momenta $k_1,k_2\ldots k_N$ scatter to
particles of momenta $ p_1,p_2\ldots p_M$ is given by
\begin{equation}
S_{k_1,\,k_2\ldots k_N \rightarrow p_1\,p_2\ldots p_M}
 =\left[\frac{\delta}{\delta a_{k_1}}\frac{\delta}{\delta a_{k_2}}\cdots 
   \frac{\delta}{\delta a_{k_N}}\frac{\delta}{\delta a_{p_1}^*}
   \frac{\delta}{\delta a_{p_2}^*} \cdots \frac{\delta}{\delta a_{p_M}^*}
  {\cal F}[\varphi]\right]_{\varphi=0}\,. 
\end{equation}
Since $F$ is not particularly 
relevant for our calculations below we can drop it in what follows. 
Further we consider $\varphi(x)$ to be an arbitrary function rather than a 
solution to $K\varphi=0$. Eventually, in obtaining the S-matrix elements we 
can choose it to be a solution to $K\varphi=0$. We thus define
\begin{eqnarray}
{\cal F}[\varphi]
       &=&\int[d\phi]\ e^{iS_0[\phi]+iS_{int}(\phi+\varphi)
	 -i\frac12\int\varphi K\varphi}\nonumber\\
       &=&\int [d\phi]\ e^{iS[\phi]-i\int\varphi K\phi}. \label{5}
\end{eqnarray}
where we added a term $\exp(-i/2 \int \varphi K \varphi )$ for 
simplifications in what follows; this of course does not contribute 
when $K\varphi =0$.
In terms of this ${\cal F}$, the S-matrix elements are given by
\begin{equation}
S_{k_1,\,k_2\ldots k_N \rightarrow p_1\,p_2\ldots p_M}
 =\left[ \alpha_{k_1}\ldots \alpha_{k_N} \alpha^*_{p_1}\ldots \alpha^*_{p_M} 
  {\cal F}[\varphi]\right]_{\varphi=0}\,. \label{6}
\end{equation}
where
\begin{equation}
\alpha_{k} =\int u_{k}(x) {\delta \over {\delta \varphi (x)}}
\end{equation} 
{}From the second of equations (\ref{5}), we see that
${\cal F}$ is the generating functional of the connected Green functions
with source $K\varphi$ and (\ref{6}) represents the well-known LSZ reduction
formula.

Consider now $\frac{\delta{\cal F}}{\delta\varphi}$. Differentiating with 
respect to $\varphi$, we find 
\begin{equation}
iK^{-1}\frac{\delta{\cal F}}{\delta\varphi}
   =\int [d\phi]~ \phi e^{iS[\phi]-i\int\varphi K\phi}.
\end{equation}
Effectively, $\phi$'s inside the integral behave as 
$iK^{-1}\frac{\delta}{\delta \varphi}$. From
(\ref{5}) we also have the equation of motion
\begin{equation}
\int[d\phi]\ \left[\frac{\delta S}{\delta \phi}
 -K\varphi\right]e^{iS[\phi]-i\int\varphi K\phi}=0,
\end{equation}
which can be written as
\begin{equation} \label{dF}
i\frac{\delta{\cal F}}{\delta\varphi}
 + \rho ({\hat \phi} ) {\cal F}-(K\varphi){\cal F}=0\,.
\end{equation}
where $\rho (\phi) =\left(\frac{\delta S_{int}}{\delta\phi}\right)$ 
and ${\hat \phi}= iK^{-1}\frac{\delta}{\delta \varphi}$.

${\cal F}$ itself generates connected as well as disconnected scattering 
processes.  If we write ${\cal F}[\varphi]=e^{iC[\varphi]}$,
where $C[\varphi]$ describes connected processes only, (\ref{dF}) reduces to
\begin{eqnarray} \label{dC}
-\frac{\delta}{\delta \varphi}{C}
  &=&K\varphi-{\cal F}^{-1}\rho ({\hat \phi}) {\cal F}\nonumber\\
  &=&K\varphi- \rho (-K^{-1}\frac{\delta C}{\delta\varphi}+{\hat \phi})\,.
\end{eqnarray}
This can be regarded as a nonlinear functional differential equation for the
S-matrix generating functional and can be used for deriving systematic
recursion rules for scattering amplitudes. We shall do this for Yang-Mills
theory in the next section.

Although it is not crucial to the discussion of recursion rules, 
one may also work with the quantum effective action or the
generating functional for one-particle irreducible vertices 
$\Gamma[\Phi]$ defined by 
\begin{equation} \label{legendre}
\Gamma[\Phi]=C[\varphi]+\int(K\varphi)\Phi\,,
\end{equation}
with the connecting relations
\begin{equation} \label{connecting}
\frac{\delta\Gamma}{\delta\Phi}=K\varphi,\qquad
-K^{-1}\frac{\delta C}{\delta\varphi}=\Phi\,.
\end{equation}
Then, (\ref{dC}) becomes
\begin{equation}
K\Phi=K\varphi-\rho (\Phi +{\hat \phi})1\,,
\end{equation}
or, in terms of $\Gamma[\Phi]$,
\begin{eqnarray} \label{dG}
\frac{\delta\Gamma}{\delta\Phi}
  &=&K\Phi+ \rho(\Phi +{\hat \phi}) 1 \nonumber\\
  &=&\left(\frac{\delta S}{\delta\phi}
     \right)_{\phi=\Phi+iK^{-1}\frac{\delta}{\delta \varphi}}1\,.
\end{eqnarray} 
The right hand side involves the derivative of $\Phi$ with respect to 
$\varphi$ which is the two-point correlator $\widetilde{G}$ given by
\begin{equation} 
\widetilde{G}(x,y)
  \equiv\left(iK^{-1}\frac{\delta}{\delta \varphi}\right)_x\Phi(y)
  =\left(iK^{-1}\frac{\delta}{\delta \varphi}
   \right)_x\left(iK^{-1}\frac{\delta}{\delta \varphi}\right)_yiC\,.
\end{equation}
It satisfies the basic Schwinger-Dyson equation for the theory, viz.,
\begin{equation} \label{SD}
\int_y\left[\frac{\delta^2\Gamma}{\delta\Phi(x)\delta\Phi(y)}
 \right]\widetilde{G}(y,z)=i\delta^{(4)}(x-z)\,.
\end{equation}
Equation (\ref{dG}) supplemented by (\ref{SD}) can thus be regarded as a 
nonlinear equation for $\Gamma[\Phi]$. From the connecting relations 
(\ref{connecting}) we find that $\frac{\delta\Gamma}{\delta\Phi}=0$ for 
$K\varphi=0$, as is appropriate for S-matrix elements. In this case,
from the definition of $\Gamma$, we have
\begin{equation}
C[\varphi] =\Gamma[\Phi]\vert_{\frac{\delta\Gamma}{\delta\Phi}=0} \,.
\end{equation}
The S-matrix is then given by
\begin{equation}
{\cal F}=\left[e^{i\Gamma[\phi]}\right]_{\frac{\delta\Gamma}{\delta\Phi}=0}\,.
\end{equation}
This relation gives a nonperturbative 
definition of the S-matrix. The free data in the solutions to 
$\frac{\delta\Gamma}{\delta\Phi}=0$
are the quantities on which ${\cal F}$ depends. 
(Perturbatively, the free data are the amplitudes $a_k,~a_k^*$ in the
solution for $\varphi$.) The fact that the S-matrix can be obtained as 
the exponential of ($i$ times ) the action evaluated on solutions of 
the equations of motion is rather well known \cite{Faddeev}, \cite{JL}.

\section{THE S-MATRIX FUNCTIONAL FOR YANG-MILLS THEORY}
We start with the gauge-fixed Lagrangian ${\cal L}$ of an 
$SU(N)$-Yang-Mills theory given by
\begin{eqnarray}
{\cal L}&=&-\frac14 F_{\mu\nu}^aF^{\mu\nu a}-\frac12(\partial\cdot A)^2
	    -\bar{c}(-\partial\cdot D)c\nonumber\\
    &=&\frac12{\rm Tr\;} F_{\mu\nu}F^{\mu\nu}+{\rm Tr\;}(\partial\cdot A)^2
	 +2{\rm Tr\;}\bar{c}(-\partial\cdot D)c\,,
\end{eqnarray}
where
\begin{eqnarray}
F_{\mu\nu}&=&F_{\mu\nu}^aT^a=\partial_\mu A_\nu-\partial_\nu A_\mu
	     +g[A_\mu,A_\nu]\,,\nonumber\\
D_\mu c&=&\partial_\mu c+g[A_\mu,c]\,,
\end{eqnarray}
with $[T^a,T^b]=f^{abc}T^c$ and ${\rm Tr\;} T^aT^b=-\frac12\delta_{ab}$. 
$T^a$ are matrices in the fundamental representation of $SU(N)$. The free part 
and the interaction parts of the action are respectively identified as
\begin{eqnarray}
{\cal L}_0&=&{\rm Tr\;} A_\mu(-\partial^2)A^\mu+2{\rm Tr\;} 
	      \bar{c}(-\partial^2)c\nonumber\\
{\cal L}^{(1)}_{int}&=&2g{\rm Tr\;}\partial_\mu A_\nu[A^\mu,A^\nu]
             +\frac{g^2}2{\rm Tr\;}[A_\mu,A_\nu]^2 \\
{\cal L}^{(2)}_{int}&=& 2g{\rm Tr\;}\partial_\mu\bar{c}[A^\mu,c]\,.
       \label{lag}
\end{eqnarray}
Following the discussion of Sec.\ 2, the functional ${\cal F}$ which gives
the S-matrix can be constructed as
\begin{equation}
{\cal F}[a]
  =\int e^{iS-i\int a_\mu^a(-\partial^2)A^{\mu a}}[dA][dc][d\bar{c}]\,,
\end{equation}
The transition amplitude for $N$ gluons of momenta $k_i$, 
polarizations $\epsilon_\mu^{(i)}$ and colors labelled
$a_1,\ldots, a_N$ to go to $M$ gluons of momenta $p_j$, polarizations
$\epsilon_\nu^{(j)}$ and color labels $a_{N+1},\ldots,a_M$ is
\begin{equation}
T[\{k_i,\epsilon_{\mu_i}^{(i)},a_i\}
    \rightarrow\{p_j,\epsilon_{\nu_j}^{(j)},a_j\}]
=\left.\int\prod_i e^{-ik_ix_i}\epsilon_{\mu_i}^{(i)}
   \frac{\delta}{\delta a_{\mu_i}^{a_i}}
\prod_je^{ip_jy_j}\epsilon_{\nu_j}^{(j)}
   \frac{\delta}{\delta a_{\nu_j}^{a_j}}{\cal F}[a]\right|_{a=0}\,.
\end{equation}

{}From (\ref{dF}) ${\cal F}$ satisfies the equation
\begin{eqnarray} \label{dF2}
-i\frac{\delta{\cal F}}{\delta a_\mu^a}
 &=& \partial^2a_\mu^a{\cal F} 
     +\int\frac{\delta S_{int}}{\delta a_\mu^a}e^{iS}\nonumber\\
 &=&\partial^2a_\mu^a{\cal F} 
     + \left.J_\mu^a(A)\right|_{A=iG\frac{\delta}{\delta a}}{\cal F}
     +\int 2g{\rm Tr\;}\partial_\mu\bar{c}[T^a,c]e^{iS}\,,
\end{eqnarray}
where
\begin{eqnarray} \label{current}
J_\mu(A(x))
 &=&J_\mu^aT^a\equiv\frac{\delta S^{(1)}_{int}}{\delta A_\mu^a}T^a\nonumber\\
 &=&-g\partial^\nu[A_\mu,A_\nu]+g[F_{\mu\nu},A^\nu]\,. 
\end{eqnarray}
and $G(x,y)=-\langle x|\partial^{-2}|y\rangle$. As we have written it, 
the term in (\ref{dF2}) involving ghosts cannot immediately be replaced 
by derivatives on ${\cal F}$. For this, we proceed as follows. Integrating
out the ghost fields in ${\cal F}$ we get Faddeev-Popov determinant 
$\det(-\partial\cdot D)=e^{{\rm tr\;}\ln(-\partial\cdot D)}$, which 
is equivalent to a term in the action $-i{{\rm tr\;}\ln(-\partial\cdot D)}$. 
This leads to a ghost-current of the form
\begin{eqnarray}
J_\mu^a(A)_{gh}
 &=&-i\frac{\delta\;{\rm tr\;}\ln(-\partial\cdot D)}{\delta A^a_{\mu}}
	    \nonumber\\
 &=&i\sum_{n=1}^\infty(-g)^{n+1}\int
    {\rm Tr\;}[T^aA_{\nu_i}(y_1)\cdots A_{\nu_n}(y_n)]_{adj}
    \partial^{\nu_1}G(x,y_1)\cdots
     \partial^{\nu_n}G(y_{n-1},y_n)\partial_\mu G(y_n,x)\,,\nonumber\\
&&
\end{eqnarray}
where we have introduced the adjoint representation of $T^a$'s by
$(T_{adj}^a)_{bc}=-f^{abc}$. (${\cal L}^{(2)}_{int}$ in this notation reads
${\cal L}^{(2)}_{int}= -g \partial_\mu \bar{c} T^a_{adj}c A^a_\mu$.)

The trace in the above equation may be written in terms of the generators in
fundamental representation using $f^{abc}=-2{\rm Tr\;} T^a[T^b,T^c]$,
\begin{equation}
{\rm Tr\;}(T^{a_1}T^{a_1}\cdots T^{a_n})_{adj}
  =2{\rm Tr\;} T^a[T^b,[T^{a_1},[T^{a_2},
    [\ldots,[T^{a_{n-1}},[T^{a_n},T^b]]\ldots]\,.
\end{equation}
Therefore  we can finally write the ghost part of the current (as a matrix 
in the fundamental representation) as
\begin{eqnarray} \label{J_ghost}
J_\mu(A)_{gh}&=&J_\mu^a(A)_{gh}T^a\\
&=&i\sum_{n=1}^\infty (-g)^{n+1}
   D_{\mu\mu_1\ldots\mu_n}(x,y_1,\ldots,y_n)
   [T^b,[A^{\mu_1}(y_1),[\ldots,[A^{\mu_{n-1}}(y_{n-1}),
   [A^{\mu_n}(y_n),T^b]]\ldots],
\nonumber
\end{eqnarray}
where 
\begin{equation}
D_{\mu\mu_1\ldots\mu_n}(x,y_1,\ldots,y_n)
 \equiv\partial_{\mu_1}G(x,y_1)\cdots
       \partial_{\mu_n}G(y_{n-1},y_n)\partial_\mu G(y_n,x).
\end{equation}

Then (\ref{dF2}) finally becomes
\begin{equation} \label{dF3}
-i\frac{\delta{\cal F}}{\delta a_\mu^a}
=\partial^2a_\mu^a{\cal F}+\left[J_\mu^a(A)
 +J_\mu^a(A)_{gh}\right]_{A=iG\frac{\delta}{\delta a}}{\cal F}\,.
\end{equation}
For connected part $C[a]=-i\ln{\cal F}[a]$,
\begin{equation} \label{dC2}
\hbar \frac{\delta C}{\delta a_\mu^a}
=\partial^2a_\mu^a+\left[J_\mu^a(A)+\hbar J_\mu^a(A)_{gh}
      \right]_{A=-\hbar G\frac{\delta C}{\delta a}
	       +i\hbar G\frac{\delta}{\delta a}}1\,.
\end{equation}

This is the basic equation for the scattering amplitudes. The
$\hbar$-expansion of this equation leads to recursion rules upto
arbitrary order. 
We have explicitly displayed the factors of $\hbar$ in (\ref{dC2}).
Recall that the interaction part of the action carries $1/\hbar$ and that
each propagator $G$ carries $\hbar$. The ghost terms, 
arising from the determinant, have $\hbar^0$. $C[a]$ in (\ref{dC2})
has an expansion of the form 
\begin{equation}
C[a]\equiv-i\ln{\cal F}[a]=\frac1\hbar C^{(0)}+C^{(1)}+\hbar C^{(2)}+\cdots\,.
\end{equation}

Starting from (\ref{dC2}) and using the above expansion for $C[a]$, 
one can, in principle, systematically derive recursion
relations to any desired order in $\hbar$.

Let us begin with a tree-level recursion formula for $C^{(0)}$. This
is simply given by 
\begin{equation}
\frac{\delta C^{(0)}}{\delta a^{\mu a}}=\partial^2a_\mu^a+J_\mu^a(A^{(0)})\,,
\end{equation}
where 
\begin{equation}
A_\mu^{(0)a}\equiv-\int G\frac{\delta C^{(0)}}{\delta a^{\mu a}}\,,
\end{equation}
i.e., 
\begin{equation} \label{tree}
\partial^2A_\mu^{(0)}=\partial^2a_\mu+J_\mu(A_\mu^{(0)})\,.
\end{equation}
This is just the equation of motion as it should be.
$A_\mu^{(0)}$ is essentially the same object as the current that 
Berends and Giele \cite{BG} used to derive tree-level recursion relation
for gluon scattering processes; if one expands $A_\mu^{(0)}$ in
powers of $a_\mu$, the coefficient function of each term gives
one-gluon off-shell current of Ref.\ \cite{BG} when multiplied by 
polarization vectors of on-shell external gluons. More generally,
if one considers $a_\mu$'s off shell, it gives the generalized current with
off-shell gluons which has been used for $q\bar q\rightarrow q\bar q gg\cdots
g$ process and some one-loop calculations \cite{MYD,M}. Also, notice that
from the form of current $J_\mu$ in (\ref{current}), 
it is obvious that color factors
are factorized from coefficient functions and we can write
\begin{equation} \label{A}
A_\mu^{(0)}
 =\sum_{n=1}^\infty\int C_{\mu\mu_1\ldots\mu_n}^{(0)}(x,y_1,\ldots,y_n)
  a^{\mu_1}(y_1)\cdots a^{\mu_n}(y_n)\,.
\end{equation}
The coefficient functions $C^{(0)}$'s do not carry color indices. We also have
$C_{\mu\mu_1}^{(0)}(x,y_1)=\delta(x-y_1)\delta_{\mu\mu_1}$.
Using (\ref{A}) in (\ref{tree}),
\begin{eqnarray}
\lefteqn{\partial^2\sum\int C_\mu^{(0)}(x,Y)a(Y)
	 -\partial^2a_\mu=J_\mu(A)}\nonumber\\
  &=&-\sum\int[2\partial^\nu C_\mu^{(0)}(x,Y)C_\nu^{(0)}(x,Z)
     -2\partial_\nu C_\mu^{(0)}(x,Z)C^{\nu (0)}(x,Y)
     +C_\nu^{(0)}(x,Y)\stackrel{\leftrightarrow}{\partial}_\mu 
      C^{\nu (0)}(x,Z)\nonumber\\
  &&\qquad\quad+C_\mu^{(0)}(Y)\partial_\nu C^{\nu (0)}(x,Z)
           -C_\mu^{(0)}(x,Z)\partial^\nu C_\nu^{(0)}(x,Y)]a(X)a(Y)\nonumber\\
    & &-\sum\int[
         2C_\nu^{(0)}(x,X)C_\mu^{(0)}(x,Y)C^{\nu (0)}(x,Z)
	 -C_\nu^{(0)}(x,X)C^{\nu (0)}(x,Y)C_\mu^{(0)}(x,Z)\nonumber\\
   &&\qquad\quad -C_\mu^{(0)}(x,X)C_\nu^{(0)}(x,Y)
      C^{\nu (0)}(x,Z)]a(X)a(Y)a(Z)\,,
\label{xrecursion}
\end{eqnarray}
where $X,\ Y$ and $Z$ are collective indices and $a(Y)$ stands for $\prod_i
a_{\mu_i}(y_i)$ etc. If $a_\mu$ is restricted to be on-shell, the terms 
containing $\partial^\nu C_\nu^{(0)}$ in the third line vanish because 
of current conservation. We can transform this equation to momentum 
space by writing
\begin{equation}
C^{(0)}_{\mu\mu_1\ldots\mu_n}(x,y_1,\ldots,y_n)
 =\int(2\pi)^4\d(k+\sum p_i)C^{(0)}_{\mu\mu_1\ldots\mu_n}(p_1,\ldots,p_n)
  e^{ik\cdot x}e^{i\sum p_i\cdot y_i}\,,
\end{equation}
where momentum conservation has been taken into account. Then 
(\ref{xrecursion}) becomes
\begin{eqnarray} \label{precursion}
P_{1,n}^2 C^{(0)}_\mu(1,\ldots,n)
&=&\sum_{m=1}^{n-1}{V_{3\mu}}^{\nu\rho}(P_{1,m},P_{m+1,n})
   C^{(0)}_\nu(1,\ldots,m)C_\rho^{(0)}(m+1,\ldots,n)\nonumber\\
&&+\sum_{m=1}^{n-2}\sum_{k=m+1}^{n-1}{V_{4\mu}}^{\nu\rho\sigma}
   C^{(0)}_\nu(1,\ldots,m)C^{(0)}_\rho(m+1,\ldots,k)
   C^{(0)}_\sigma(k+1,\ldots,n)\,,\nonumber\\
&&
\end{eqnarray}
where $P_{i,j}=p_i+\cdots+p_j$ and $V_3^{\mu\nu\rho}$, 
$V_4^{\mu\nu\rho\sigma}$ are color-ordered vertices,
\begin{eqnarray} \label{vertices}
V_3^{\mu\nu\rho}(p,q)&=&-g\{g^{\nu\rho}(p-q)^\mu+g^{\rho\mu}(2q^\nu+p^\nu)
		  -g^{\mu\nu}(2p^\rho+q^\rho)\}\,,\nonumber\\
V_4^{\mu\nu\rho\sigma}&=&-g(2g^{\mu\rho}g^{\nu\sigma}
	    -g^{\mu\nu}g^{\rho\sigma}-g^{\mu\sigma}g^{\nu\sigma})\,.
\end{eqnarray}
This equation is the recursion relation
for currents with off-shell gluons derived in \cite{BG},\cite{MYD}.

In deriving higher order recursion relations for S-matrix elements, 
we must correct (\ref{dC2}) by including the appropriate renormalization 
constants. 
As usual, we interpret fields and coupling constants as renormalized ones and
assume that renormalization counterterms are included in ${\cal L}_{int}$
of (\ref{lag}). Explicitly,
\begin{eqnarray}
{\cal L}_{int}
  &=&-\frac12\delta Z_3{\rm Tr\;}(\partial_\mu A_\nu-\partial_\nu A_\mu)^2
     -2gZ_1{\rm Tr\;}\partial_\mu A_\nu[A_\mu,A_\nu] 
     -\frac{g^2}2Z_1^2Z_3^{-1}{\rm Tr\;}[A_\mu,A_\nu]^2\nonumber\\
  && -2\delta {\tilde Z_3}{\rm Tr\;}\bar{c}(-\partial^2)c
    -2gZ_1{\tilde Z_3}/{Z_3}{\rm Tr\;}\partial_\mu\bar{c}[A_\mu,c]\,,
\end{eqnarray}
where $\sqrt{Z_3}A_\mu$, $\sqrt{\tilde Z_3}c$ and $gZ_1Z_3^{-3/2}$ are bare 
quantities with $\delta Z_3=Z_3-1, \delta {\tilde Z_3}={\tilde Z_3}-1$.
Correspondingly, $J_\mu$ becomes
\begin{eqnarray}
J_\mu(A)&=&-g\partial_\nu[A_\mu,A_\nu]+g[F_{\mu\nu},A_\nu]\nonumber\\
  &&  -\frac12\delta Z_3\partial_\nu(\partial_\mu A_\nu-\partial_\nu A_\mu)
  -g(Z_1-1)(\partial_\nu[A_\mu,A_\nu]-[F_{\mu\nu},A_\nu])\nonumber\\
  && +g^2Z_1(Z_1/Z_3 -1)[[A_\mu,A_\nu],A_\nu]\,.
\end{eqnarray}
The extra counterterms contribute to one-loop or higher orders and will cancel
ultraviolet divergences from loop integrals. In addition,
to account for the freedom of arbitrary finite renormalization 
we include a finite wavefunction renormalization constant $z_3$ in 
the external field, i.e., 
$a_\mu\rightarrow\tilde{a}_\mu\equiv a_\mu/\sqrt{z_3}$.

Now we are ready to discuss one-loop recursion relations.
Keeping terms relevant upto one-loop order,
\begin{eqnarray}
A_\mu&=& -G\frac{\delta C^{(0)}}{\delta a^\mu}
      -\left(G\frac{\delta C^{(1)}}{\delta a^\mu}-\frac12\delta z_3
	G\frac{\delta C^{(0)}}{\delta a^\mu}\right)+\cdots\nonumber\\
&=&A_\mu^{(0)}+A_\mu^{(1)}+\cdots\,
\end{eqnarray}
The current can similarly be expanded as
$J_\mu(A)=J_\mu^{(0)}+J_\mu^{(1)}+\cdots $,
with
\begin{eqnarray}
J_\mu^{(0)}&=&-g\partial^\nu[A^{(0)}_\mu,A^{(0)}_\nu]
	      +g[F_{\mu\nu}(A^{(0)}),A^{(0)\nu}]\,,\nonumber\\
J_\mu^{(1)}
 &=&-g\partial^\nu([A^{(0)}_\mu,A^{(1)}_\nu]+[A^{(1)}_\mu,A^{(0)}_\nu])
   +g[F_{\mu\nu}(A^{(0)}),A^{(1)\nu}]
   +g[D_\mu A^{(1)}_\nu-D_\nu A^{(1)}_\mu,A^{(0)\nu}]\nonumber\\
 &&-\delta Z_1 J_\mu^{(0)}(A^{(0)})
   -\frac{\delta Z_3}2\partial^\nu
    (\partial_\mu A^{(0)}_\nu-\partial_\nu A^{(0)}_\mu) \nonumber\\
 &&+g^2(\delta Z_1-\delta Z_3)[[A^{(0)}_\mu,A^{(0)}_\nu],A^{(0)\nu}]\,,
\end{eqnarray}
where $D_\mu\equiv\partial_\mu+g[A^{(0)}_\mu,\ ]$.
In (\ref{dC2}) we need
\begin{eqnarray}
J_\mu(A+iG\frac{\delta}{\delta a})1
&=&J_\mu(A) -g\partial^\nu[iG\frac{\delta}{\delta a^\mu},A^{(0)}_\nu]
  +g[i\partial_\mu G\frac{\delta}{\delta a^\nu}
  -i\partial^\nu G\frac{\delta}{\delta a^\mu},A^{(0)}_\nu]\nonumber\\
&&-g^2\left([[iG\frac{\delta}{\delta a^\mu},A^{(0)}_\nu],A^{(0)\nu}]
  +[[A^{(0)}_\mu,iG\frac{\delta}{\delta a_\nu}],A^{(0)}_\nu] \right)+\cdots\,.
\end{eqnarray}  
Upto the one-loop order, the functional
derivative term in the ghost current has no
contribution,
\begin{equation}
J_\mu(A+iG\frac{\delta}{\delta a})_{gh}1=J_\mu(A^{(0)})_{gh}+\cdots\,.
\end{equation}
Collecting all these, we get an equation for $A^{(1)}_\mu$,
\begin{eqnarray} \label{oneloop}
\partial^2A^{(1)}_\mu
&=&J_\mu^{(1)}+J_\mu(A^{(0)})_{gh}\nonumber\\
 && -g\partial^\nu[iG\frac{\delta}{\delta a^\mu},A^{(0)}_\nu]
  +g[i\partial_\mu G\frac{\delta}{\delta a^\nu}-i\partial^\nu 
   G\frac{\delta}{\delta a^\mu},A^{(0)}_\nu]\nonumber\\
&&-g^2\left([[iG\frac{\delta}{\delta a^\mu},A^{(0)}_\nu],A^{(0)\nu}]
  +[[A^{(0)}_\mu,iG\frac{\delta}{\delta a_\nu}],A^{(0)}_\nu] \right)\,,
\end{eqnarray}
where $A^{(0)}_\mu$ is the solution of (\ref{tree}). Notice that, in this 
case, $iG\frac{\delta}{\delta a_\mu}$-terms contract color indices when 
acting on $A^{(0)}_\mu$ and so the color decomposition does not occur as 
in tree-level case. However, it is possible to write $A^{(1)}_\mu$ as 
a sum of color-factorized amplitudes and the proof has been given 
using string-theory argument \cite{BK1} and color flow diagrams\cite{BK2}. 
Here we give a simple proof based on
(\ref{oneloop}).

First we study the action of $iG\frac{\delta}{\delta a_\mu}$ on 
$A^{(0)}_\mu$. From (\ref{A}),
\begin{eqnarray}
[G\frac{\delta}{\delta a^\mu},A^{(0)}_\nu]
  &=&\sum_{n=1}^\infty\sum_{m=1}^n\int 
   C_{\nu\nu_1\ldots\nu_{m-1}\mu\nu_{m+1}\ldots\nu_n}^{(0)}
      (x,y_1,\ldots,y_{m-1},y,y_{m+1},\ldots,y_n)G(x,y)\nonumber\\
  &&\hspace{15mm}\times [T^a,a_{\nu_1}(y_1)\cdots a_{\nu_{m-1}}(y_{m-1})T^a
        a_{\nu_{m+1}}(y_{m+1})\cdots a_{\nu_n}(y_n)]\,.
\end{eqnarray}
With the help of Fierz identity for $SU(N)$
\begin{equation} \label{fiertz}
(T^aXT^a)_{ij}=-\frac12\left(\delta_{ij}{\rm Tr\;} X-\frac1N X_{ij}\right)\,,
\end{equation}
it becomes
\begin{equation} \label{gda}
[G\frac{\delta}{\delta a^\mu},A^{(0)}_\nu]
  =\sum_{n=1}^\infty\sum_{m=0}^n\int 
   \widetilde{C}_{\nu\mu}(x,1,\ldots,m;m+1\ldots,n)
   a(1)\cdots a(m){\rm Tr\;}[a(m+1)\cdots a(n)]\,,
\end{equation}
where
\begin{eqnarray}
\widetilde{C}_{\nu\mu}(x,1,\ldots,m;m+1\ldots,n)
 &=&-\frac12\int_y G(x,y)
  [C_{\nu\nu_1\ldots\nu_m\mu\nu_{m+1}\ldots\nu_n}^{(0)}
      (x,y_1,\ldots,y_m,y,y_{m+1},\ldots,y_n)  \nonumber\\
  &&\hspace{15mm} -C_{\nu\nu_m\ldots\nu_n\mu1\ldots\nu_m}^{(0)}
                     (x,y_{m+1},\ldots,y_n,y,y_1,\ldots,y_m)]\,.
\end{eqnarray}
(Here the $1/N$-term in (\ref{fiertz}) does not contribute because of the
commutator; notice also that the summation in (\ref{gda}) starts with $m=0$.) 
Thus differentiation of
$A_\mu^{(0)a}$ produces terms with trace over substrings of $a_\mu$'s. Also,
from (\ref{J_ghost}), we see that the ghost current has the same structure,
\begin{eqnarray}
J_\mu(A^{(0)})_{gh}
 &=&i\sum_{n=1}^\infty (-g)^{n+1} D_\mu(x,1,\ldots,n)\nonumber\\
 &&\times \sum_{k=0}^n\sum_{\{i_l\}}
  \{T^bA^{(0)}(i_1)\cdots A^{(0)}(i_k)T^bA^{(0)}(i_{k+1})
  \cdots A^{(0)}(i_n)\nonumber\\
 &&\hspace{15mm}-A^{(0)}(i_1)\cdots A^{(0)}(i_k)T^bA^{(0)}(i_{k+1})
		\cdots A^{(0)}(i_n)T^b\},
\end{eqnarray}
where $\sum_{\{i_l\}}$ is the sum over all permutations of $\{1,\ldots,n\}$ 
such that $i_1<\ldots<i_k$ and $i_{k+1}>\ldots>i_n$. Using (\ref{fiertz}), we 
get
\begin{eqnarray} \label{J_ghost2}
J_\mu(A^{(0)})_{gh}
&=&-\frac{i}2\sum_{n=1}^\infty \sum_{k=0}^n
   \sum_{\sigma\in S_{n;k}}g^{n+1} (-1)^k
   \widetilde{D}_\mu(x,\sigma_1,\ldots,\sigma_n)\nonumber\\
&&\times A^{(0)}(1)\cdots A^{(0)}(k){\rm Tr\;}[A^{(0)}(k+1)\cdots A^{(0)}(n)],
\end{eqnarray}
where $S_{n;k}$ is the set of all permutations of $\{1,2,\ldots,n\}$ that
preserves the ordering of $\{\alpha\}\equiv\{1,\ldots,k\}$ and the cyclic 
ordering of $\{\beta\}\equiv\{n,n-1,\ldots,k+1\}$, while allowing for all 
possible relative orderings between elements in the two sets (For example, 
$(1,n,2,\ldots,k,n-1,\ldots)$ is in $S_{n;k}$ but 
$(2,1,\ldots,k,n,n-1,\ldots)$ is not.); $\widetilde{D}_\mu$ is given by
\begin{equation}
\widetilde{D}_\mu(x,1,\ldots,n)
  =D_\mu(x,1,\ldots,n) -(-1)^nD_\mu(x,n,n-1,\ldots,1).
\end{equation}
Since (\ref{oneloop}) is at most linear in $A^{(1)}_\mu$, 
it is clear that $A^{(1)}_\mu$ can be written as 
\begin{equation}  \label{A1}
A^{(1)}_\mu(x)
  =\sum_{n=1}^\infty\sum_{m=0}^n\int
   C_{m\,\mu}^{(1)}(x,1,\ldots,m;m+1,\ldots,n)
      a(1)\cdots a(m){\rm Tr\;}[a(m+1)\cdots a(n)]\,.
\end{equation}
Then in (\ref{oneloop}) we can separately equate terms with the same trace
structure and $C_{m\,\mu}^{(1)}$'s with different $m$'s do not mix
with each other. The functions we need are then
$C^{(1)}_{n\,\mu}=C^{(1)}(x,1,\ldots,n)$ corresponding to the term in 
(\ref{A1}) with
no trace. These obey the recursion rule obtained by substituting 
$A^{(1)}=\sum_{n=1}^{\infty}\int C^{(1)}(x,1,\ldots,n)a(1)\cdots a(n)$ in
(\ref{oneloop}). The other amplitudes can be obtained from $C^{(1)}_{n\,\mu}$.

The coefficient functions corresponding to different trace structures have
simple relations among them (which allow us to construct the amplitudes with
subtraces from $C^{(1)}_{n\,\mu}$). 
We have \cite{BK1},\cite{BK3}
\begin{equation} \label{cnck}
C_{m\,\mu}^{(1)}(x,1,\ldots,m;m+1,\ldots,n)=(-1)^{n-m}
 \sum_{\sigma\in S_{n;m}}C_{n\,\mu}^{(1)}(x,\sigma_1,\ldots,\sigma_n)\,,
\end{equation}
For this, we show that the right hand side of (\ref{cnck}) satisfies the
same equation as that for the left hand side. 
This is most easily seen for the ghost current which we shall consider first. 
Obviously, an equation like (\ref{cnck}) holds for $\widetilde{D}_\mu$'s
(with the identification of the summation indices $k$ in 
(\ref{J_ghost2}) and $m$ in (\ref{cnck})), if 
$A^{(0)}_\mu(x)=\int C^{(0)}_\mu(x,X)a(X)$ in (\ref{J_ghost2}) is replaced 
by its lowest order term $a_\mu$. 
For the terms with more than one $a_\mu$'s from one $A^{(0)}$, there are
potential discrepancies between both sides. This is because in (\ref{cnck})
the sumation is over all $\sigma\in S_{n;m}$ while, in (\ref{J_ghost2}), we sum
only over $\sigma\in S_{n;k}$ ($k<m$) which does not include such permutations 
that mix indices from both $\{\alpha\}$ set and $\{\beta\}$ set within one tree 
structure. However, the extra terms in (\ref{cnck}) cancel out thanks to the
symmetry properties of color-ordered vertices,
\begin{eqnarray} \label{symmetry}
&&V_3^{\mu\nu\rho}(p,q)+V_3^{\mu\rho\nu}(q,p)=0\,,\nonumber\\
&&V_4^{\mu\nu\rho\sigma}+V_4^{\mu\rho\nu\sigma}+V_4^{\mu\sigma\nu\rho}=0\,,
\end{eqnarray}
which, when applied to (\ref{precursion}) together with induction, leads to 
identities for $1\le m<n$,
\begin{equation} \label{identity}
\sum_{\sigma\in S'_{n;m}}C^{(0)}_\mu(\sigma_1,\ldots,\sigma_n)=0\,,
\end{equation}
where $S'_{n;m}$ is the same as $S_{n;m}$ defined above except that it does 
not include the cyclic permutations of $\{\beta\}$ set.
[This equation can be considered as a generalization of the so-called
``dual Ward identity'' for tree amplitudes which corresponds to $m=1$ case
\cite{KLN},\cite{DY}.] Thus (\ref{cnck}) holds for ghost current.

We shall now show that the relation (\ref{cnck}) connecting amplitudes with
different trace structures holds for the nonghost terms in (\ref{A1}) as well.
Towards this, consider the terms with differentiation 
$G\frac{\delta}{\delta a}$ in 
(\ref{oneloop}). Diagramatically, those terms correspond to one-loop gluon
diagrams with the external leg $x$ directly connected to the loop. In our
setting, we can proceed as follows. Applying (\ref{precursion}) repeatedly
to $C^{(0)}_\mu(x,1,\ldots,m,y,m+1,\ldots,n)$ we can identify the
internal line for each term from (\ref{precursion}).
It connects $x$ and $y$ and the other external legs are ordered in clockwise
direction in such a way that $1,\ldots,m$-th legs are below the line while
$m+1,\ldots,n$-th legs are above the line (Fig.\ 1). Then we draw a line 
connecting $x$ and $y$ which enclose $m+1,\ldots,n$-th legs so that it
represents that the legs inside the loop are traced. Then, it is easy 
to see which terms generate which color structures. Obviously, given such
a diagram, we get the same subtrace structure from diagrams made by altering
the relative order of trees belonging to different sets while keeping the
order of $\{1,\ldots,m\}$ and the cyclic order of $\{m+1,\ldots,n\}$ (Fig.\
2). Moreover, they are trivially related to terms which contribute to 
$C^{(1)}_{n\,\mu}$, i.e., we can simply pull out the trees inside the loop to
outside with minus sign, using symmetry property of 
vertices (\ref{symmetry}). Notice that during this procedure the order of
$\{m+1,\ldots,n\}$ is also reversed. Thus, essentially we have the same 
situation as in the case of ghost current and the relation of the type 
(\ref{cnck}) holds in this case as well.

Now it remains to consider the term from $J^{(1)}$ which contains 
$C^{(1)}_\mu$'s with less number of legs. It corresponds to the gluon-loop
diagram with the external leg $x$ attached to a tree. This case, however, is
not much different from the previous one and we can argue in the same way with
the help of (\ref{identity}) if needed. Finally, there are counterterm
contributions for $C^{(1)}_{n\,\mu}(x,1,\ldots,n)$ case in contrast to
$C^{(1)}_{m\,\mu}(x,1,\ldots,m;m+1,\ldots,n)$ $(m<n)$. But again the identity
(\ref{identity}) guarantees that those terms cancel out when the summation in
(\ref{cnck}) is done. This completes the proof that both the left and the
right hand sides of (\ref{cnck}) satisfy the same recursion relation. Since
(\ref{cnck}) trivially holds for $n=3$, they are indeed equal to each other.

Beyond one-loop, it is clear from (\ref{dC2}) that $A_\mu^{(k)}$ will
in general have terms with $k$ subtraces,
\begin{eqnarray}
A_\mu^{(k)}(x)
 &=&\sum_{n=1}^\infty\sum_{\{m_l\}}\int
   C_{m_1m_2\ldots m_k\mu}^{(k)}(x,1,\ldots,m_1;m_1+1,\ldots,m_2;\ldots;
     m_k+1\ldots n)\nonumber\\
&&\hspace{20mm}\times a(1)\cdots a(m_1) {\rm Tr\;}[a(m_1+1)\cdots 
  a(m_2)]\cdots{\rm Tr\;}[a(m_k+1)\cdots a(n)]\,,\nonumber\\
&&
\end{eqnarray}
and each $C_{m_1m_2\ldots m_k\mu}^{(k)}$ with different $k$ will satisfy
its own equation. It might be possible to find simple relations between them
as in one-loop case.

\vspace{10mm}

CK was supported in part by the Korea Science and Engineering
Foundation and VPN was supported in part by the National Science Foundation
Grant Number PHY-9322591.

\newpage

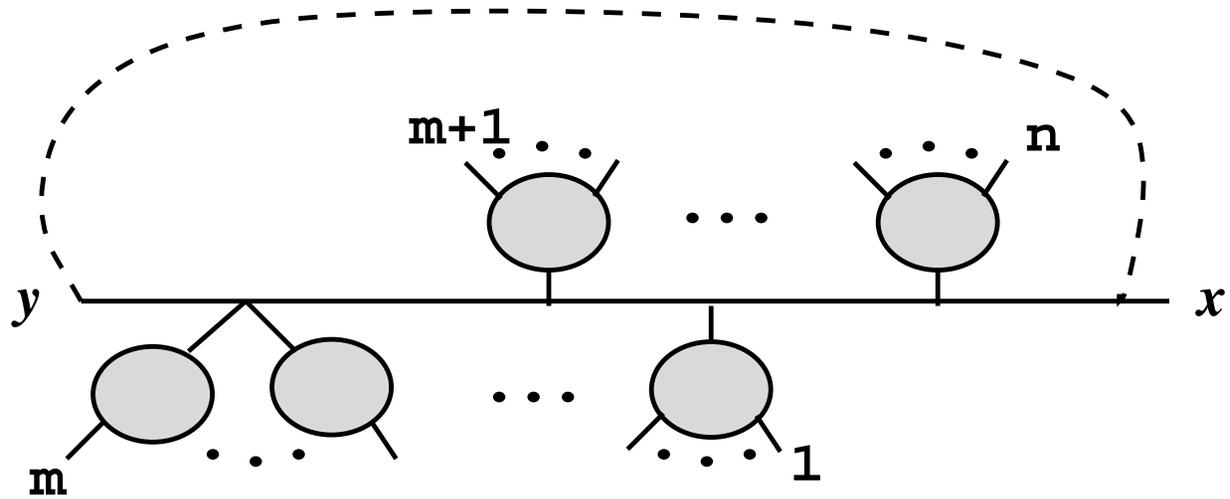
\begin{figure}
\caption{A typical term in $C^{(0)}_\mu(x,1,\ldots,m,y,m+1,\ldots,n)$.
$y$ is connected to $x$ and the loop encloses legs $m+1,\ldots,n$ which are
under a trace.}
\end{figure}
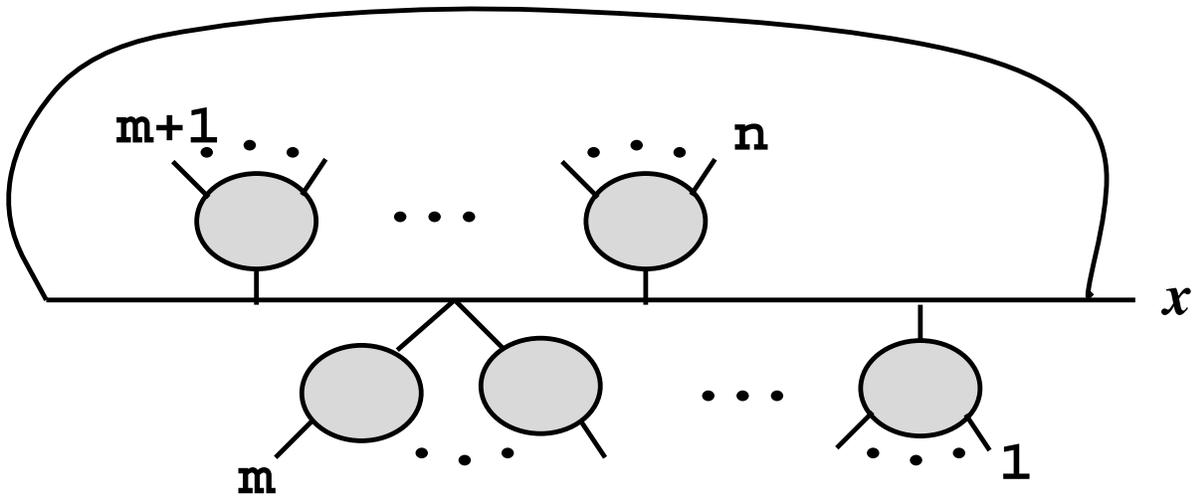
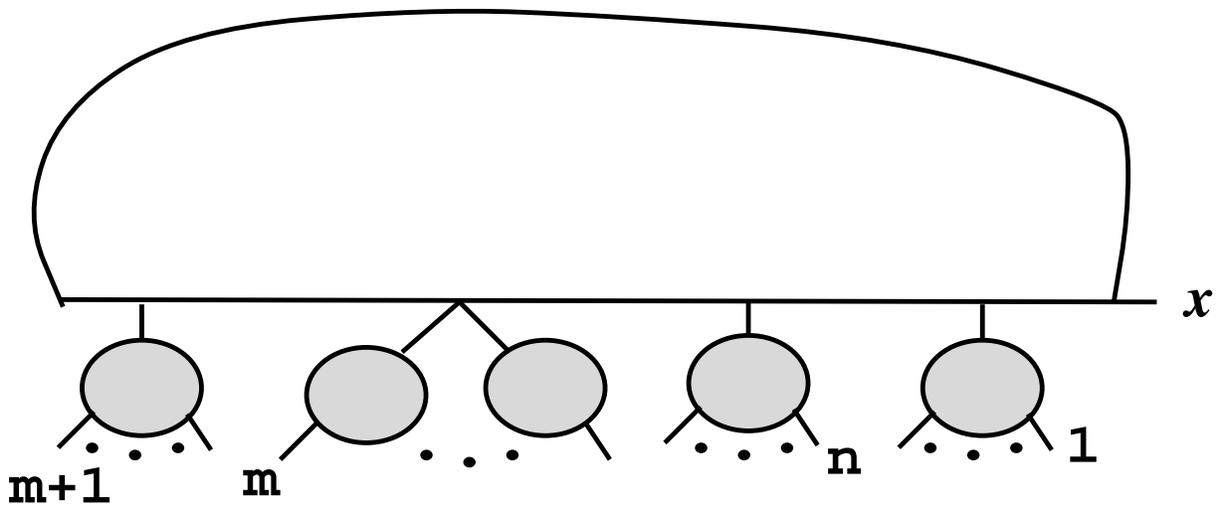
\begin{figure}
\caption{(a)A diagram obtained by changing the relative order of traced legs 
and the others from Fig.\ 1. It contributes to
$C_{m\,\mu}^{(1)}(x,1,\ldots,m;m+1,\ldots,n)$; 
(b) A diagram obtained by pulling the legs out of the loop in Fig.\ 2(a). 
It contributes to $C_{n\,\mu}^{(1)}(x,1,\ldots,n)$.}
\end{figure}
\end{document}